\documentclass[12pt,cite,epsfig,amssymb]{article}
\pdfoutput=1
\usepackage[english]{babel}
\usepackage{epsfig}
\usepackage{amssymb}
\usepackage{amsmath}
\usepackage{cite}
\usepackage{authblk}
\usepackage{wrapfig}

\begin{document}

\title{On the peculiar features of the instantaneous angular distribution of synchrotron radiation}

\author{V. G. Bagrov$^{1,2}$\thanks{bagrov@phys.tsu.ru}, A. N. Kasatkina$^{1}$
\thanks{ane4ka.com.ru@sibmail.com}, A. A. Pecheritsyn$^{1}$\thanks{pecher@phys.tsu.ru}}

\affil{$^{1}$Tomsk State University, 634050, Tomsk, Lenin Ave, 36, Russia}
\affil{$^{2}$Institune of High Current Electronics, SB RAS, \\
634055, Tomsk, Akademichesky Ave, 4, Russia}
\maketitle

\begin{abstract}
The instantaneous angular distribution of synchrotron radiation is investigated. The radiation space is divided into two parts. The first part of the space is the interior of a cone with the apex located at the point of a radiating charge, the angular span being $2\alpha_0$ and the central axis oriented along the instantaneous velocity of the charge. The second part of the space is the complement of the first part in the entire space.

It is shown that the radiation in the ultrarelativistic limit is entirely contained in the first part of the space, and the radiation in the second part (with a non-zero $\alpha_0$) vanishes.\\

\end{abstract}

\section{Introduction}

A theoretical study of angular distributions of the power of synchrotron radiation (SR) is sufficiently well developed and presented, for example, in \cite{Bagrov-1965, Sokolov_et_al-1966, Sokolov-1968, SR-1999, Rad}. However, some features of SR angular distributions have not been considered and may prove to be interesting
not only theoretically, but also experimentally feasible.

In this paper, we consider the instantaneous spatial distribution of SR power. This distribution was first studied in \cite{Bagrov-1965}. In particular, in \cite{Bagrov-1965} it was established that the SR instantaneous distribution in the relativistic case is concentrated along the velocity vector of a radiating charge (electron). We propose here a method of dividing the space into two parts, such that in the ultrarelativistic case all the radiation power concentrates in the first part of the space, and in the second part the radiation tends to zero.

\section{Spatial structure of the SR instantaneous angular distribution}

The spatial structure of the SR instantaneous angular distribution can be specified in the following coordinates (also indicated in \cite{Bagrov-1965}). The origin of the coordinate system is chosen at the 
\begin{wrapfigure}[12]{l}{0.3\linewidth}
%\begin{figure}[bt]
\centering
\includegraphics[width=5cm]{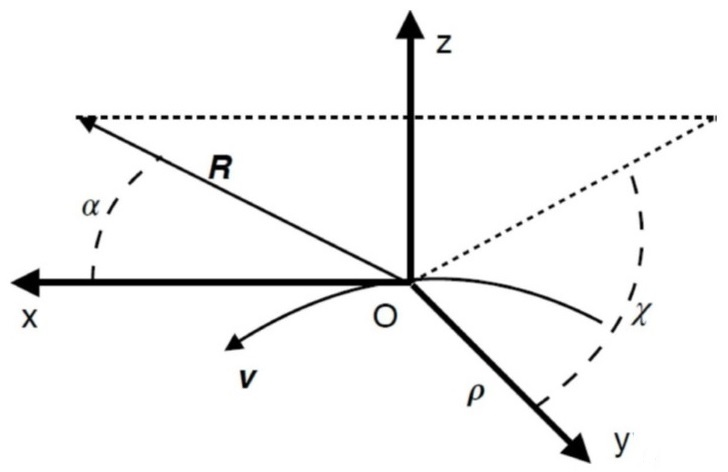} 
\parbox[t]{5cm}{\caption{Coordinate system.}
\label{fig:coord}}
%\end{figure}
\end{wrapfigure}
point where the radiating charge is located. The $x$-axis is directed along the electron velocity, the $y$-
axis is directed toward the center of a circular trajectory, and the $z$-axis is chosen so that the coordinate system is right-handed (when the charge moves in a constant and uniform magnetic field, the $z$-axis is parallel to this external magnetic field). The radius of the circular orbit of the radiating particle is denoted by $\rho$; in our coordinate system it is oriented along the $y$-axis. The angle between the $x$-axis and the vector $\textbf{R}$ is denoted by $\alpha$ $(0 \leqslant \alpha \leqslant \pi)$, while the angle between the projection of the vector $\textbf{R}$ onto the $yz$-plane and the $y$-axis is denoted by $\chi$ $(0 \leqslant \chi < 2\pi)$. The coordinate system is illustrated by Fig.~\ref{fig:coord}.

Under these assumptions, the instantaneous angular distribution has the form [1-5]
\begin{equation}\label{b.1}
d W = \frac{e^2 \omega_c^2 \beta^2 (1 - \beta^2)}{4\pi c} \frac{[(\beta - \cos\alpha)^2 + (1 - \beta^2) \sin^2\alpha \sin^2\chi]}{(1 - \beta \cos\alpha)^5} \sin\alpha\,d\alpha \,d\chi \, .
\end{equation}
The following notation is introduced: $e$ is the charge value; $c$ is the speed of light; $v = c\beta$ is the speed of a radiating particle $(0 \leqslant \beta < 1)$; $\omega_c = (|e H|)/(m_0 c)$ is the cyclotron frequency; $m_0$ is the rest mass of the particle; $H$ is the magnetic field strength. The radius of the orbit $\rho$ of the radiating particle is related with the cyclotron frequency and the velocity by the formula
\begin{equation}\notag
\rho = \sqrt{\frac{\beta^2}{1 - \beta^2}}\, \frac{c}{\omega_c}\,.
\end{equation}

It is formula (\ref{b.1}) that determines (see also \cite{Bagrov-1965}) the fact that the instantaneous angular distribution in the relativistic case is mainly concentrated in a narrow cone with the central axis coinciding in direction with the instantaneous particle velocity $v$. The angular span
of this cone is $\sim \gamma^{- 1} = \sqrt{1 - \beta^2}$.

Integration over the angular variables in (\ref{b.1}) leads to well-known expression for full radiation power \cite{Bagrov-1965, Sokolov_et_al-1966, Sokolov-1968, SR-1999, Rad}
\begin{equation}\label{b.2}
W = \frac{2 e^2 \omega^2_c \beta^2}{3 c (1 - \beta^2)} = \frac{2 e^2 \omega^2_c}{3 c} (\gamma^2 - 1)\,.
\end{equation}

The entire radiation space in question will be represented as the sum of two subspaces. The first subspace is the interior of a circular cone emerging from the origin of coordinates, with the angular span being $2 \alpha_0 \, (0<\alpha_0 \leqslant \pi)$ and the central axis coinciding with the charge velocity ($x$-axis). The second subspace is the remaining part of the space with the first subspace removed. The radiation power in the first subspace can be obtained from expression (\ref{b.1}) by integrating over $\chi$ within $0 \leqslant \chi < 2\pi$ and by integrating over $\alpha$ within $\alpha_0 \leqslant \alpha \leqslant \pi$. In the second subspace, the radiation power is given by integrating over $\chi$ within $0 \leqslant \chi < 2\pi$ and by integrating over $\alpha$ within $\alpha_0 \leqslant \alpha \leqslant \pi$. It is essential to notice the value $0 < \alpha_0 \leqslant \pi$, meaning that $\alpha_0$ is larger than zero. The radiation power in the first subspace will be denoted by $W_1$, and the radiation power in the second subspace is denoted by $W_2$. Next, we introduce the notation
\begin{equation}\notag
W_1 = \frac{e^2 \omega_c^2}{c} P(\beta, \alpha_0), \ \ W_2 = \frac{e^2 \omega_c^2}{c} G(\beta, \alpha_0), \end{equation}
where  $P(\beta, \alpha_0), G(\beta, \alpha_0)$ are dimensionless functions.
The equality $W = W_1 + W_2$ and formula (\ref{b.2}) imply the relation
\begin{equation}\label{b.3}
P(\beta, \alpha_0) +  G(\beta, \alpha_0) = \frac{2 \beta^2}{3(1 - \beta^2)} \rightarrow   P(\beta, \alpha_0) = \frac{2 \beta^2}{3(1 - \beta^2)} -  G(\beta, \alpha_0) .
\end{equation}
Expression (\ref{b.3}) allows one to obtain the function $P(\beta, \alpha_0)$ once the function $G(\beta, \alpha_0)$ is known.

Using expression ((\ref{b.1}) and making a trivial integration over the variable $\chi$, we get the following integral representation
for the function $G(\beta, \alpha_0)$:
\begin{equation}\label{b.4}
G(\beta, \alpha_0) = \beta^2 (1 - \beta^2) \int^\pi_{\alpha_0} \frac{2(\beta - \cos\alpha)^2 + (1 - \beta^2) \sin^2\alpha} {4(1 - \beta \cos\alpha)^5} \sin\alpha\,d\alpha \, .
\end{equation}
Integration over the variable $\alpha$ in (\ref{b.4}) does not pose a problem, and the final expression for $G(\beta, \alpha_0)$ has the form
\begin{equation}\label{b.5}
G(\beta, \alpha_0) = \frac{\beta^2 (1 - \beta^2)(1 + q)\Gamma (\beta; q)}{48 (1 + \beta)^2 (1 - \beta q)^4 } \,; \ \ q = \cos\alpha_0 , \ \ 1 > q \geqslant - 1;
\end{equation}
where it has been denoted
\begin{equation}\label{b.6}
\Gamma (\beta; q) = 3\beta^3(1 - q)(1 + 3q^2) + 2\beta^2(3 - 13q + 19q^2 - q^3) + \beta(11 - 51q + 9q^2 - q^3) + 4(4 - q + q^2) \,.
\end{equation}

For derivatives with respect to $\beta$, it is easy to obtain the formulas
\begin{equation}\label{b.7}
G'(\beta, \alpha_0) = \frac{\beta (1 + q) M(\beta, q)}{48 (1 + \beta)^2 (1 - \beta q)^5} , \ \ P'(\beta, \alpha_0) = \frac{4 \beta}{3(1 - \beta^2)^2} -  G'(\beta, \alpha_0) ,
\end{equation}
where $M(\beta, q)$ stands to denote
\begin{equation}\notag
M(\beta, q) = 2 [1 + \beta (q - 1) + \beta^2 (q - 1) - \beta^3 q]\Gamma (\beta; q) + \beta (1 - \beta q - \beta^2  + \beta^3 q) \Gamma' (\beta; q) ,
\end{equation}
\begin{equation}\label{b.8}
\Gamma' (\beta; q) = 9\beta^2(1 - q + 3q^2 - 3q^3) + 4\beta(3 - 13q + 19q^2 - q^3) + 11 - 51q + 9q^2 - q^3 .
\end{equation}

\section{Properties of the functions $P(\beta, \alpha_0)$ and $G(\beta, \alpha_0)$}

Here, we discuss the simplest properties of the function $G(\beta, \alpha_0)$ defined by expressions (\ref{b.5}) and (\ref{b.6}). As has been already noted, the value of $q < 1$ is strictly less than the unity, therefore the value of $G(\beta, \alpha_0)$ with permissible parameters is a finite and infinitely differentiable function with respect to the variables $\beta, q$ and admits a continuous transition $\beta \rightarrow 1$ with $G(1, \alpha_0) = 0$. As it follows from (\ref{b.5}), we also have $G(0, \alpha_0) = 0$ for $\beta = 0$. Thus, considering the function $G(\beta, \alpha_0)$ as a function of $\beta$ on the closed segment $0 \leqslant \beta \leqslant 1$ with a fixed $q = \cos\alpha_0$, we can see that this function is continuous (and infinitely differentiable) on the indicated segment and is equal to zero at the ends. It is easy to find the first nonzero terms at the ends of this segment:
\begin{equation}\notag
G(\beta \sim 1, \alpha_0) \approx (1 - \beta^2) \frac{(1 + q)(3 - q)}{16 (1 - q)^2} , \ \ G(\beta \sim 0, \alpha_0) \approx \beta^2 \frac{(1 + q)[15 + (1 - 2q)^2]}{48} .
\end{equation}

It is obvious that at an inner point there is a maximum of the function $G(\beta, \alpha_0)$, which is found, according to formulas (\ref{b.7}) and (\ref{b.8}), from the equation
\begin{equation}\label{b.9}
M(\beta, q) = 0.
\end{equation}
Thus, in the second part of the space for a given $\alpha_0$ there exists an upper limit of the radiated power.

The function $P(\beta, \alpha_0)$ is a monotonically increasing (up to infinity, with $\beta \rightarrow 1$) function of $\beta$ for any given
$q = \cos\alpha_0$. Therefore, it is only $P(\beta, \alpha_0)$ that determines the SR power in the ultrarelativistic limit.

\begin{figure}[bt]
\centering
\includegraphics[width=14cm]{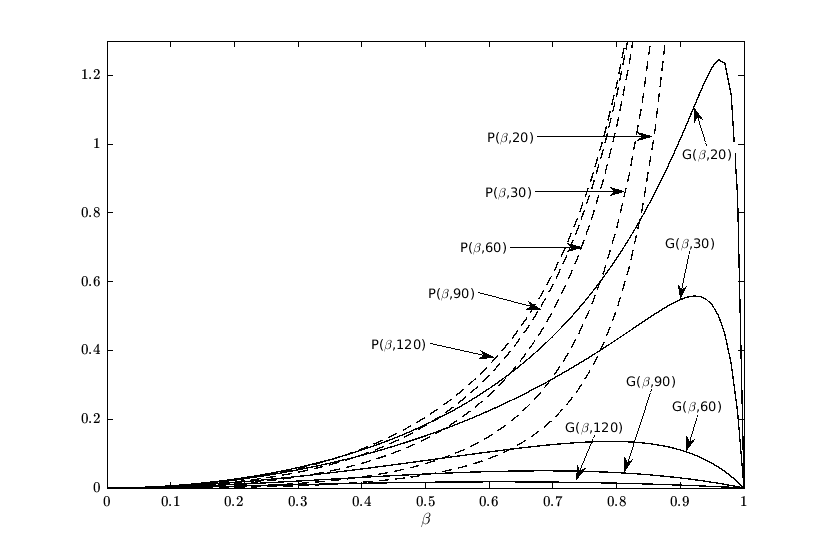} 
\parbox[t]{9cm}{\caption{Functions $G(\beta,\alpha_0)$ and $P(\beta,\alpha_0)$.}
\label{fig:GP}}
\end{figure}

For illustration, we plot in Fig.~\ref{fig:GP} the graphs of the functions $P(\beta, \alpha_0)$ and $G(\beta, \alpha_0)$ for some values of $\alpha_0$. The behavior of the graphs confirms the above reasoning. \\

The numerical results of calculations for the functions $G(\beta, \alpha_0)$ and $P(\beta, \alpha_0)$ are given in the Table~\ref{table:GP}, where, for the corresponding $\alpha_0$, in the first line we display the function $G(\beta, \alpha_0)$, and in the second line, the function $P(\beta, \alpha_0)$.\\

\begin{figure}[bt]
\centering
\includegraphics[width=14cm]{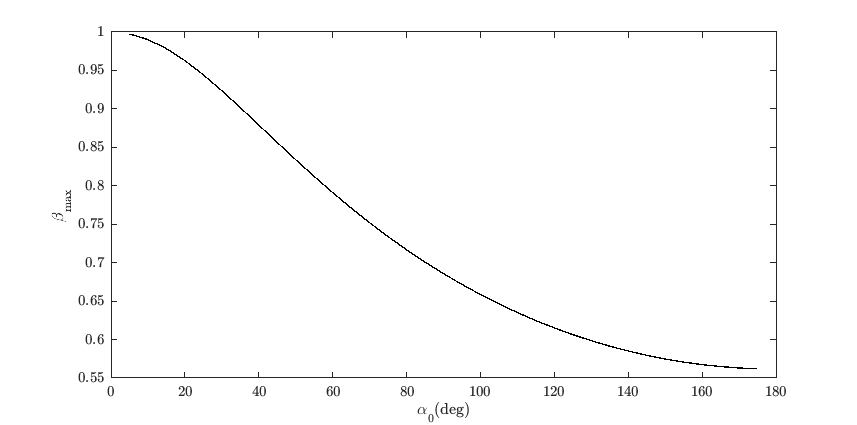} 
\parbox[t]{9cm}{\caption{Dependence  $\beta_{\mbox{max}}$ from $\alpha_0$.}
\label{fig:maxg}}
\end{figure}

In Fig.~\ref{fig:maxg}, we also present a plot for the maximum of the function $G(\beta, \alpha_0)$ as a solution of equation (\ref{b.9}). \\

\begin{table}[thp]
\caption{Function values $G(\beta,\alpha_0)$ and $P(\beta,\alpha_0)$.}\label{table:GP}
\begin{tabular}{|r|r|r|r|r|r|r|r|r|r|r|}
\hline
    & \multicolumn{9}{c|}{$\beta$} &  \\
\cline{2-10}
 $\alpha_0$   &  0.1   &  0.2   &  0.3   &  0.4   &  0.5   &  0.6   &  0.7   &  0.8   &  0.9   &
$G_{\mbox{max}}$    \\
\hline
10 & 0.0066 & 0.0272 & 0.0642 & 0.1224 & 0.2113 & 0.3493 & 0.5765 & 0.9971 & 1.9856 & 4.9530 \\
\cline{2-10}
 & 0.0001 & 0.0006 & 0.0018 & 0.0046 & 0.0109 & 0.0257 & 0.0641 & 0.1881 & 0.8565 & \\
\hline
20 & 0.0063 & 0.0257 & 0.0594 & 0.1104 & 0.1841 & 0.2892 & 0.4413 & 0.6676 & 1.0147 & 1.2476 \\
\cline{2-10}
 & 0.0004 & 0.0021 & 0.0066 & 0.0166 & 0.0381 & 0.0858 & 0.1993 & 0.5176 & 1.8274 & \\
\hline
30 & 0.0059 & 0.0234 & 0.0528 & 0.0950 & 0.1516 & 0.2250 & 0.3177 & 0.4313 & 0.5477 & 0.5584 \\
\cline{2-10}
 & 0.0008 & 0.0043 & 0.0131 & 0.0320 & 0.0706 & 0.1500 & 0.3228 & 0.7539 & 2.2944 & \\
\hline
40 & 0.0054 & 0.0209 & 0.0458 & 0.0797 & 0.1223 & 0.1730 & 0.2303 & 0.2882 & 0.3110 & 0.3145 \\
\cline{2-10}
 & 0.0013 & 0.0069 & 0.0201 & 0.0473 & 0.1000 & 0.2020 & 0.4102 & 0.8970 & 2.5312 & \\
\hline
50 & 0.0049 & 0.0184 & 0.0393 & 0.0664 & 0.0986 & 0.1342 & 0.1700 & 0.1967 & 0.1824 & 0.1994 \\
\cline{2-10}
 & 0.0019 & 0.0094 & 0.0267 & 0.0606 & 0.1236 & 0.2408 & 0.4705 & 0.9885 & 2.6597 & \\
\hline
60 & 0.0043 & 0.0160 & 0.0335 & 0.0553 & 0.0799 & 0.1049 & 0.1263 & 0.1352 & 0.1106 & 0.1353 \\
\cline{2-10}
 & 0.0024 & 0.0117 & 0.0324 & 0.0716 & 0.1424 & 0.2701 & 0.5143 & 1.0500 & 2.7315 & \\
\hline
70 & 0.0038 & 0.0140 & 0.0286 & 0.0460 & 0.0646 & 0.0816 & 0.0934 & 0.0933 & 0.0693 & 0.0954 \\
\cline{2-10}
 & 0.0029 & 0.0138 & 0.0374 & 0.0809 & 0.1577 & 0.2934 & 0.5471 & 1.0919 & 2.7728 & \\
\hline
80 & 0.0034 & 0.0120 & 0.0242 & 0.0380 & 0.0516 & 0.0629 & 0.0686 & 0.0646 & 0.0446 & 0.0688 \\
\cline{2-10}
 & 0.0034 & 0.0157 & 0.0418 & 0.0890 & 0.1706 & 0.3121 & 0.5719 & 1.1206 & 2.7975 & \\
\hline
90 & 0.0029 & 0.0103 & 0.0201 & 0.0308 & 0.0406 & 0.0476 & 0.0498 & 0.0447 & 0.0293 & 0.0499 \\
\cline{2-10}
 & 0.0038 & 0.0175 & 0.0458 & 0.0962 & 0.1816 & 0.3274 & 0.5907 & 1.1405 & 2.8128 & \\
\hline
100 & 0.0025 & 0.0085 & 0.0163 & 0.0243 & 0.0311 & 0.0354 & 0.0357 & 0.0308 & 0.0194 & 0.0361 \\
\cline{2-10}
 & 0.0042 & 0.0192 & 0.0496 & 0.1026 & 0.1911 & 0.3396 & 0.6048 & 1.1544 & 2.8227 & \\
\hline
110 & 0.0021 & 0.0069 & 0.0129 & 0.0186 & 0.0232 & 0.0256 & 0.0251 & 0.0210 & 0.0128 & 0.0258 \\
\cline{2-10}
 & 0.0047 & 0.0209 & 0.0531 & 0.1084 & 0.1991 & 0.3494 & 0.6155 & 1.1642 & 2.8293 & \\
\hline
120 & 0.0016 & 0.0053 & 0.0097 & 0.0137 & 0.0166 & 0.0178 & 0.0171 & 0.0140 & 0.0083 & 0.0179 \\
\cline{2-10}
 & 0.0051 & 0.0225 & 0.0563 & 0.1133 & 0.2056 & 0.3572 & 0.6234 & 1.1712 & 2.8338 & \\
\hline
130 & 0.0012 & 0.0038 & 0.0068 & 0.0095 & 0.0112 & 0.0119 & 0.0111 & 0.0090 & 0.0053 & 0.0119 \\
\cline{2-10}
 & 0.0055 & 0.0239 & 0.0591 & 0.1175 & 0.2110 & 0.3631 & 0.6294 & 1.1762 & 2.8369 & \\
\hline
140 & 0.0008 & 0.0025 & 0.0044 & 0.0060 & 0.0070 & 0.0073 & 0.0068 & 0.0054 & 0.0031 & 0.0073 \\
\cline{2-10}
 & 0.0059 & 0.0252 & 0.0615 & 0.1210 & 0.2152 & 0.3677 & 0.6337 & 1.1798 & 2.8390 & \\
\hline
150 & 0.0005 & 0.0015 & 0.0025 & 0.0034 & 0.0039 & 0.0040 & 0.0037 & 0.0029 & 0.0017 & 0.0040 \\
 \cline{2-10}
& 0.0063 & 0.0263 & 0.0634 & 0.1236 & 0.2183 & 0.3710 & 0.6369 & 1.1823 & 2.8405 & \\
\hline
160 & 0.0002 & 0.0007 & 0.0011 & 0.0015 & 0.0017 & 0.0017 & 0.0016 & 0.0012 & 0.0007 & 0.0018 \\
\cline{2-10}
 & 0.0065 & 0.0271 & 0.0648 & 0.1255 & 0.2205 & 0.3733 & 0.6389 & 1.1839 & 2.8414 & \\
\hline
170 & 0.0001 & 0.0002 & 0.0003 & 0.0004 & 0.0004 & 0.0004 & 0.0004 & 0.0003 & 0.0002 & 0.0004 \\
\cline{2-10}
 & 0.0067 & 0.0276 & 0.0657 & 0.1266 & 0.2218 & 0.3746 & 0.6401 & 1.1849 & 2.8419 & \\
\hline
\end{tabular}
\end{table}

\section{Conclusion}

The main conclusion of this paper is the following statement. Dividing the space $SR$ into the two parts described in the work, we distinguish the first subspace in which the velocity vector of a radiating particle is located and the second subspace in which this vector is absent. Considering the radiation power in the second subspace as a function of the particle velocity, we can see that the radiation power
initially increases, reaches a maximum, then it begins to decrease with a growing speed and vanishes in the ultrarelativistic case. The fact that the radiation vanishes in the ultrarelativistic case is quite interesting physically. Thus, in the ultrarelativistic case, all the radiation is concentrated in the first subspace, regardless of its size (however, it is essential that this size be non-zero).

One can give another physical interpretation of the result obtained here. Considering the angular distribution of the instantaneous power of synchrotron radiation as a function of the energy of a radiating particle, we are aware that most of the power is concentrated,
with an increase of the energy, in the vicinity of a line parallel to the instantaneous charge velocity. However, this concentration is accompanied by the fact that radiation along the other lines begins to decrease with an increase of the energy and then vanishes. In the ultrarelativistic limit, it is only the increasing part of radiation (see expression (\ref{b.3})) that determines its character.

The question arises of an experimental confirmation for the results obtained. Since in accelerators and capacitors the electron rotates with a high frequency $\omega = \sqrt{1 - \beta^2} \omega_c$, it is only time-averaging characteristics altering the instantaneous distribution that can be observed. The instantaneous distribution can be observed if the temporal resolution of experimental installations (for modern accelerators) is of the order of $10^{- 13}$ sec., which is expected to be achieved in the near future.

\section*{Funding informations}

This work was supported in part by RFBR grant No. 18-02-
00149 and by the Program for Improving TSU’s Competitiveness among World Leading Scientific and Educational
Centers.


\begin{thebibliography}{9}

\bibitem{Bagrov-1965} V. G. Bagrov. 
 Angular Distribution of Radiation from a Charge in an External 
 Field in Accordance with Classical Theory.
 Soviet Journal of Optics and Spectroscopy. 
 1965, v. 18, N. 4, p. 541-544. 

\bibitem{Sokolov_et_al-1966} A. A. Sokolov, I. M. Ternov, V. G. Bagrov. 
 The Classical Theory of the Synchrotron Radiation.
 In: The Synchrotron Radiation. Ed. A. A. Sokolov, I. M. Ternov.\\
 Moscow: Nauka, 1966. P. 18-71.  (In Russian). 

\bibitem{Sokolov-1968} A. A. Sokolov, I. M. Ternov.
 Synchrotron Radiation. 
 Academie - Verlag, Berlin. 1968. P. 202.

 \bibitem{SR-1999} Synchrotron Radiation Theory And Its Development.
 Ed. V. A. Bordovitsyn.
 World Scientific. Singapore$\bullet$New Jersey$\bullet$London. 1999. - 447 p.

\bibitem{Rad} Radiation Theory of Relativistic Particles.
Editor: V. A. Bordovitsyn.
Moscow, Fizmatlit. 2002. 576 p. ISBN 5-9221-0258-3.

\end{thebibliography}
\end{document}